\documentclass[aps,pre,reprint,superscriptaddress,showpacs,preprintnumbers,amsmath,amssymb,nofootinbib,longbibliography]{revtex4-1}
\usepackage[dvipdfmx]{graphicx}
\usepackage{dcolumn}   
\usepackage{bm}        
\usepackage{amssymb, amsmath, amsfonts}   
\usepackage{lipsum}        
\usepackage{color}
\DeclareGraphicsExtensions{{.pdf}}
\begin{document}
\newcommand{\noter}[1]{{\color{red}{#1}}}
\newcommand{\noteb}[1]{{\color{blue}{#1}}}
\newcommand{\field}{\left( \boldsymbol{r}\right)}
\newcommand{\paren}[1]{\left({#1}\right)}
\newcommand{\vect}[1]{\boldsymbol{#1}}
\newcommand{\uvect}[1]{\tilde{\boldsymbol{#1}}}
\newcommand{\vdot}[1]{\dot{\boldsymbol{#1}}}
\newcommand{\vder}{\boldsymbol{\nabla}}
%
\widetext
%
\title{
Localized and extended dynamical correlation lengths in jammed packings of soft athermal disks under slow shear}
\author{Shoichi Takahata}
\thanks{These two authors contributed equally}
\affiliation{Department of Physics, Tohoku University, Aoba, Aramaki, Aoba-ku, Sendai 980-8578, Japan}
\affiliation{Mathematics for Advanced Materials-OIL, AIST, Sendai 980-8577, Japan}
\author{Norihiro Oyama}
\thanks{These two authors contributed equally}
\affiliation{Graduate School of Arts and Sciences, The University of Tokyo, Tokyo 153-8902, Japan}
\affiliation{Mathematics for Advanced Materials-OIL, AIST, Sendai 980-8577, Japan}
\author{Kuniyasu Saitoh}
\affiliation{Department of Physics, Faculty of Science, Kyoto Sangyo University, Motoyama, Kamigamo, Kita-ku, Kyoto 603-8555, Japan}
\date{\today}
\begin{abstract}
Dynamics of jammed packings of soft athermal disks under finite-rate shear are studied by means of molecular dynamics simulations.
Particularly, we investigate the spatial structures of stress drop events, which are expected to provide information about plasticity.
Investigating the displacement fields during stress drop events, we show that there are qualitatively different two types of events in the low rate limit: localized ones and extended ones.
We further investigate the time evolution of events and clarify that both types of events are due to oscillatory motion of the stress, which is unique for systems under finite-rate shear.
The difference between two types of events is the regime that events reside in: while localized events take place during plastic events, extended ones occur in the elastic branch.
\end{abstract}
\maketitle
%
\section{Introduction}
Soft athermal particles such as suspensions, emulsions, foams, and granular materials are ubiquitous in nature \cite{review0}
and a better understanding of their mechanical properties is crucial to engineering and science \cite{lemaitre,alexander}.
When the soft athermal particles are densely packed, the system exhibits the \emph{jamming transition} \cite{gn0,gn1,gn2,gn3,gn4}
such that the particles are rigid against global deformations \cite{rs0,rs1}.
Continuously shearing the dense packing of the particles, the system experiences the macroscopic yielding \cite{yield_t0,yield_t1,yield_t2,yield_t3,yield_t4} and then reaches a steady state \cite{review-rheol0,review-rheol1}.
In the steady state, the injection of energy (by shear) is dissipated not only by inelastic interactions between the particles \cite{energydissipation} but also by plastic events, or topological rearrangements of particles.
Because the plastic events release the elastic energy stored in the system, the shear stress keeps (linearly) increasing and (suddenly) decreasing in the steady state.
If the rate of global deformations is vanishingly small, the plastic events are intermittent \cite{aval_hatano}.
The statistical properties of these intermittent plastic events (known as \emph{avalanches}) has been the focus of modern non-equilibrium statistical mechanics \cite{dahmen0,aval_mean1,aval_mean2}.

Microscopic structures of the particle rearrangements under shear have been extensively studied by
experiments of e.g.\ granular materials \cite{corl0,corl9,Combe} and colloidal glasses \cite{pdf2,corl4,corl5,corl6}
as well as by molecular dynamics (MD) simulations \cite{pdf0,spectrum,saitoh11,saitoh12,dh_shear0,corl2,corl3}.
In these studies, large scale collective motions of \emph{non-affine displacements} were commonly observed.
Interestingly, spatial structures of non-affine displacements are ``vortex-like" and have been closely analyzed by
e.g.\ probability distribution functions (PDFs) \cite{pdf0,pdf1,pdf2}, spatial correlation functions \cite{corl0,corl9,corl4,corl5,corl6,corl2,corl3}, energy spectrum \cite{spectrum,corl8}, and hydrodynamic modes \cite{saitoh11}.
The collective motions of non-affine displacements are pronounced if the system is dense and driven by a small shear rate \cite{saitoh12}.
Then, they are often called \emph{rigid clusters} \cite{dh_shear0} which play a key role in the rheology \cite{rheol0} and self-diffusion \cite{diff_shear_md0,diff_shear_md1,saitoh15} of soft athermal particles under shear.
The typical size of collective motions, or rigid clusters, is also dependent on the distance from jamming;
it diverges with the increase of particle density below jamming \cite{rheol0} and spans the whole system and is proportional to the linear system size above jamming \cite{pdf1}.
Note that the linear scaling of the size (above jamming) was also found in athermal quasi-static (AQS) simulations \cite{corl2,corl3} and thus is not the effect of finite shear rate.
Furthermore, the size of collective motions is also considered as the limit of continuum descriptions of amorphous solids \cite{rs0,rs1}.
For instance, Tanguy et al. had revealed that their mechanical responses to global deformations are well predicted by elasticity theory
if the length scale is greater than the spatial correlations between non-affine displacements \cite{tanguy0,tanguy1,tanguy2,tanguy3}.
Because their numerical analyses are based on sound characteristics (i.e.\ dispersion relations and attenuation coefficients) of the particles,
one can expect that the large scale collective motions has a link to sound excitation under shear.

Plastic events have attracted a particular interest~\cite{ReviewBarrat} since they are the cause of complex rheological behavior or the stationarity under shear~\cite{MaloneyPRE2006}.
Now it is widely-accepted that displacement field with localized
quadrupolar patterns, or the so-called shear transformations (STs), are the elementary processes of plastic deformations~\cite{STZs-1,*STZs-2}.
Since STs are surrounded by long-ranged elastic field, the corresponding vibrational eigenmodes are sometimes referred to as quasilocalized modes~\cite{Mizuno2017PNAS,Shimada2018PRE,LernerPRE2020}.
On the other hand, ST events themselves can be viewed as singularities (force dipoles) embedded in an elastic body~\cite{PicardEPJE2004} and in this sense, they can be viewed as localized structures in nature.
We note that, still, the characteristic length can become system-spanning when multiple STs form system-spanning avalanches~\cite{LemaitrePRL2009, LinPNAS2014, Oyama2020INM}.  
These knowledge have been gained mainly by AQS
simulations where the thermal fluctuations are ignored and shear is
imposed quasistatically~\cite{ReviewBarrat, Oyama2020avalanche}.
It is important to mention that while plastic events are detected in a well-defined manner by intermittent stress (or potential energy) drop events under AQS condition~\cite{Oyama2020avalanche,Zhang2017PRE}, such events do not necessarily represent plasticity under finite-rate shear~\cite{TsamadosEPJE2010,Oyama2020INM}.

In this paper, we numerically investigate the elastoplasticity of dense packings of soft athermal particles under finite-rate shear.
We focus on the system above jamming and study how the particle rearrangements during \emph{plastic events} are affected by the finite rate of simple shear deformations.
We first characterize the shear rate-dependence of the displacement fields during events by the participation ratio, a quantitative measure of collectivity.
In the low shear-rate limit, surprisingly, the probability distribution function (PDF) of the participation ratio becomes bimodal, although the PDF under AQS shear is known to exhibits a unimodal shape~\cite{BaileyPRL2007}.
Analysis on the spatial structure of them clarifies that those peaks correspond to qualitatively different two types of typical events: extended ones and localized ones.
We further discuss the possible physical origins of these two types of events by careful observations of time evolution of events.
%

In the following, we introduce our numerical methods in Sec.\ \ref{sec:methods} and show our results in Sec.\ \ref{sec:results}.
In Sec.\ \ref{sec:discussion}, we discuss the connection between conventional (system spanning) correlation lengths and our new length scale.
All the technical details of our numerical analyses are summarized in Appendixes \ref{ap:Ndep} and \ref{ap:clusters}.
%
\section{Numerical methods}
\label{sec:methods}
We employ molecular dynamics (MD) simulations of soft athermal particles in two dimensions \cite{spectrum,saitoh11,saitoh12,saitoh14}.
To avoid crystallization of the system, we prepare a 50:50 binary mixture of $N$ particles,
where different kinds of particles have the same mass $m$ and different diameters, $d$ and $1.4 d$ \cite{gn0,gn1}.
The force between the particles, $i$ and $j$, in contact is modeled by a ``linear spring-dashpot" \cite{dem},\
i.e.\ $\bm{f}_{ij}=(k\xi_{ij}-\eta \dot{\xi}_{ij})\bm{n}_{ij}$, with the stiffness $k$ and viscosity coefficient $\eta$.
The force is parallel to the normal unit vector $\bm{n}_{ij}=\bm{r}_{ij}/|\bm{r}_{ij}|$,
where $\bm{r}_{ij}\equiv\bm{r}_i-\bm{r}_j$ with the particle positions, $\bm{r}_i$ and $\bm{r}_j$, represents the relative position.
In addition, $\xi_{ij}=R_i+R_j-|\bm{r}_{ij}|>0$ is the overlap between the particles, and $\dot{\xi}_{ij}$ is its time derivative,
where $R_i$ ($R_j$) is the radius of particle $i$ ($j$).
The stiffness and viscosity coefficient determine the time scale as $t_0\equiv\eta/k$
and are adjusted such that the \emph{normal restitution coefficient} of the particles is given by
$e=\text{exp}(-\pi/\sqrt{2mk/\eta^2-1})\approx 0.87$ \cite{dem}.

We randomly distribute the $N$ particles in an $L\times L$ square periodic box and relax the system to a mechanically stable state \cite{FIRE}.
The packing fraction of the particles is given by $\phi=0.88$
which is much higher than the \emph{jamming transition density} $\phi_J\simeq 0.8433$ \cite{gn3,gn4,rheol0}.
Then, we apply simple shear deformations to the system under the Lees-Edwards boundary conditions \cite{lees}.
In each time step, we apply affine deformation to the system by replacing every particle position $\bm{r}_i=(x_i,y_i)$ with $(x_i+\Delta\gamma y_i,y_i)$ ($i=1,\dots,N$)
and then numerically integrate equations of motion of the particles with a small time increment $\Delta t$ \cite{spectrum,saitoh11,saitoh12,saitoh14}.
Here, $\Delta\gamma$ is the strain increment so that the shear rate is defined as $\dot{\gamma}\equiv\Delta\gamma/\Delta t$.
Note that, to control the shear rate, we change both $\Delta\gamma$ and $\Delta t$ within the constraints,\
$\Delta \gamma\le 10^{-7}$ and $\Delta t \le 0.1 t_0$.

In our MD simulations, we use the system sizes (the number of particles), $N=8192$, $16384$, and $32768$, and vary the scaled shear rate in the range $10^{-8}\le\dot{\gamma}t_0\le 10^{-3}$.
In addition, we only analyze the data in a steady state, where the applied strain exceeds unity, $\gamma>1$.
%
\section{Results}
\label{sec:results}
In this section, we introduce the stress drop events (Sec.\ \ref{sec:plastic})
and measure participation ratios for displacement fields during those events (Sec.\ \ref{sec:bimodal}).
We analyze spatial structures of the displacement fields (Sec.\ \ref{sec:spatial})
and quantify their microscopic collectivity by extracting coherent clusters from the system (Sec.\ \ref{sec:corre_len}).
%
\subsection{Stress drop events}
\label{sec:plastic}
\begin{figure}
\includegraphics[width=\linewidth]{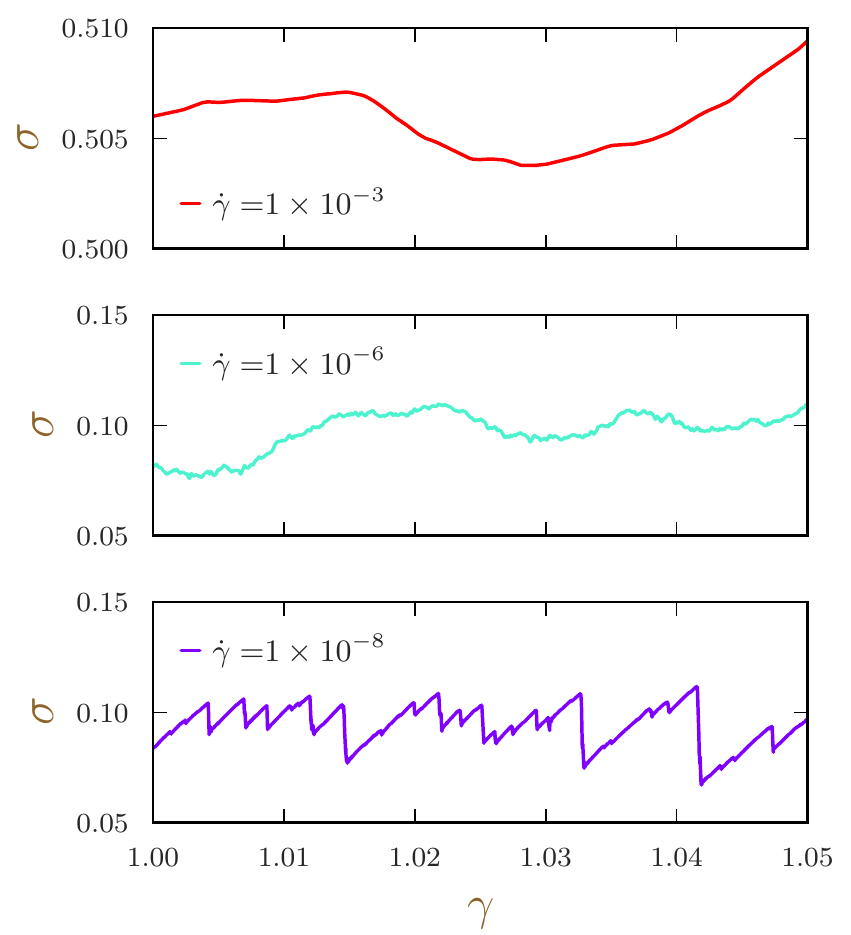}
\caption{
\emph{Stress-strain curves}, $\sigma$ vs.\ $\gamma$, where the scaled shear rate decreases from $\dot{\gamma}t_0=10^{-3}$ to $10^{-8}$ (from top to bottom) as shown in the legends.
The system size is $N=16384$.
\label{fig:s_s}}
\end{figure}
Under AQS shear, we can detect plastic events without any ambiguities by stress (or potential energy) drop events~\cite{Oyama2020avalanche,Zhang2017PRE}.
However, under finite-rate shear, we cannot directly connect stress drop events with plasticity~\cite{TsamadosEPJE2010,Oyama2020INM}.
Still, numerically~\cite{HatanoSciRep2015,BaresPRE2017} and experimentally~\cite{Antonaglia2014PRL, DenisovSciRep2017,DenisovNatCommun2016,BaresPRE2017}, stress (or potential energy) drop events have been used as the indicator of plastic events and their statistical properties have been studied from the perspective of the avalanche criticality.
Following those previous works, we employ the stress drop events.
The shear stress in our system is defined as follows:
\begin{equation}
    \sigma = -\frac{1}{L^2}\sum_{i,j}f_{ijx}^\mathrm{el}r_{ijy}~.
    \label{eq:sigma}
\end{equation}
Here, $f_{ijx}^\mathrm{el}=k\xi_{ij}n_{ijx}$ is the $x$-component of the elastic force and $r_{ijy}$ is the $y$-component of the relative position $\bm{r}_{ij}$ between the particles, $i$ and $j$, in contact (see Sec.\ \ref{sec:methods}).
We do not include the viscous force $f_{ijx}^\mathrm{vis}=-\eta\dot{\xi}_{ij}n_{ijx}$ and kinetic contribution in the stress.
Figure \ref{fig:s_s} displays \emph{stress-strain curves},\ i.e.\ $\sigma$ vs.\ $\gamma$, in a steady state ($\gamma>1$),
where we use the system size $N=16384$ and decrease the scaled shear rate from $\dot{\gamma}t_0=10^{-3}$ to $10^{-8}$ (from top to bottom).
As can be seen, the shapes of stress-strain curves are strongly dependent on $\dot{\gamma}$:
If the shear rate is large ($\dot{\gamma}t_0=10^{-3}$), the shear stress $\sigma$ smoothly changes around its mean value.
However, decreasing the shear rate, we observe that the curve becomes jerky ($\dot{\gamma}t_0=10^{-6}$)
and exhibits a serrated pattern if the shear rate is sufficiently small ($\dot{\gamma}t_0=10^{-8}$).
Note that such a serrated pattern is typical of the results of AQS simulations \cite{aval_quasi0}.

In a steady state, the shear stress $\sigma$ fluctuates around its mean value (Fig.\ \ref{fig:s_s}).
We can detect a \emph{stress drop event} such that the derivative of $\sigma$ with respect to $\gamma$ is negative \cite{aval_hatano},\ i.e.\
\begin{equation}
\frac{d\sigma}{d\gamma} < 0~.
\label{eq:drop}
\end{equation}
The $k$th stress drop event starts at $\gamma_{Ck}$ and stops at $\gamma_{Ck}+n_k\Delta\gamma$, with $n_k$ being a positive integer that describes the \emph{avalanche duration}.
Note that we expect $n_k>1$ in most cases.
\begin{figure}
\includegraphics[width=\linewidth]{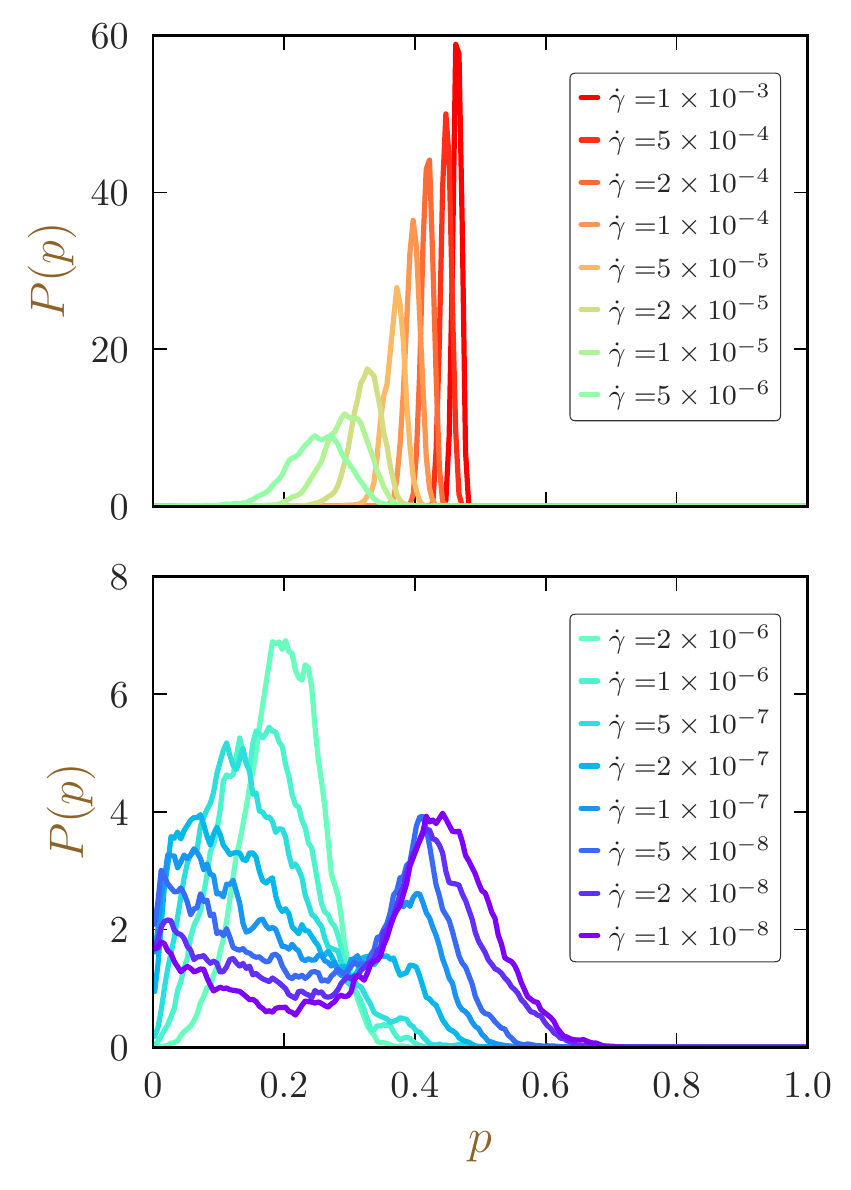}
\caption{
PDFs of participation ratios, $P(p)$, for large shear rates (top) and small shear rates (bottom).
Different colors represent different values of the scaled shear rate $\dot{\gamma}t_0$ (as listed in the legends), where the system size is $N=16384$.
Note that the different scale is used for vertical axes.
\label{fig:PDF_p}}
\end{figure}
\begin{figure*}
\includegraphics[width=\linewidth]{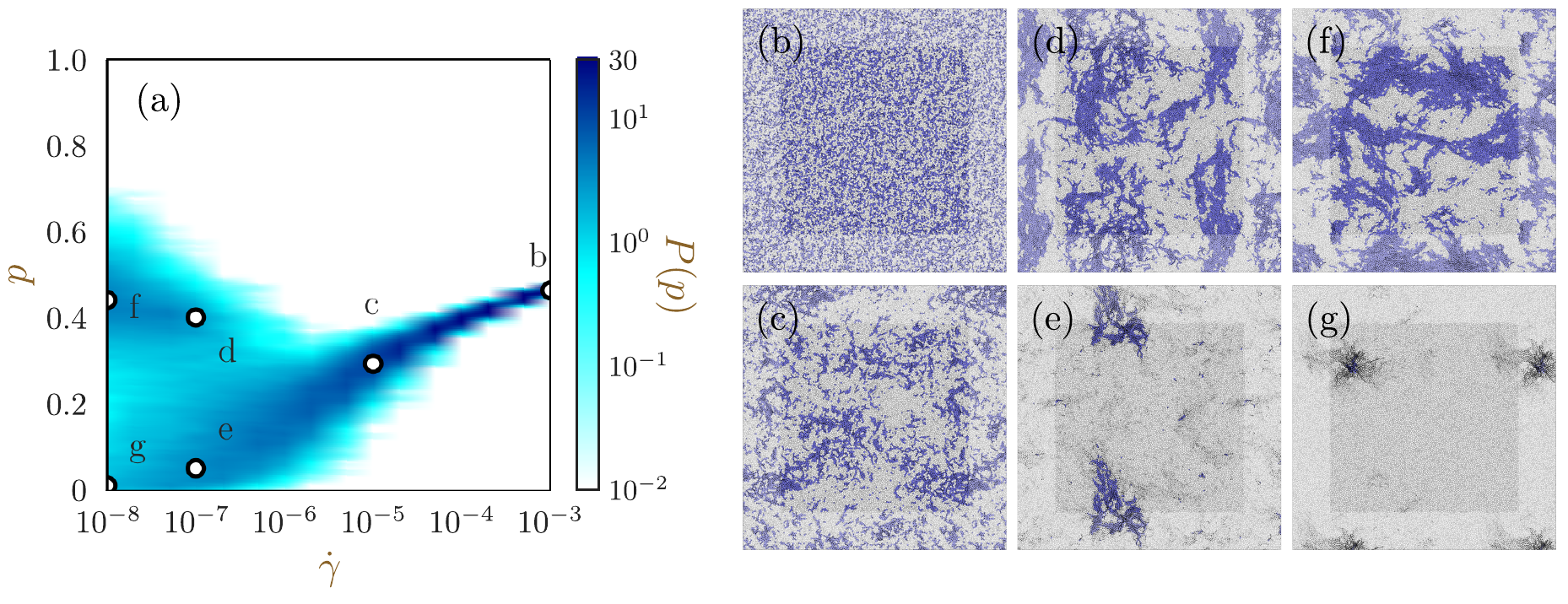}
\caption{
(a) An intensity plot of the PDFs of participation ratios, $P(p)$ (gray scale), with different values of the scaled shear rate $\dot{\gamma}t_0$ (horizontal axis).
Note that the PDF is normalized for each $\dot{\gamma}$.
(b-g): Spatial distributions of the non-affine displacements (arrows) during a stress drop event, where the corresponding values of $p$ and $\dot{\gamma}$ for (b-g) are indicated in (a).
The \emph{mobile particles} (see Sec.\ \ref{sec:corre_len}) are highlighted in blue and the non-affine displacements are properly normalized to increase the visibility (readers are encouraged to zoom in to access the fine resolution images).
Copied images due to the periodic boundary conditions are shown in slightly lighter colors.
The system size is as in Fig.\ \ref{fig:s_s}.
\label{fig:vis}}
\end{figure*}
%
\subsection{Participation ratios}
\label{sec:bimodal}
To characterize the shear-rate dependence of nonaffine displacements during stress drop events, we introduce the \emph{participation ratio} of displacement fields as \cite{aval_quasi1}
\begin{equation}
    p_k = \frac{\left(\sum_{i=1}^N d_{i,k}^2\right)^2}{N\sum_{i=1}^N d_{i,k}^4}~,
    \label{eq:p}
\end{equation}
where $d_{i,k}$ is the magnitude of the nonaffine displacement of the particle $i$ during the $k$th stress drop event.
The participation ratio is unity, $p_k=1$, if every magnitude of the non-affine displacements is the same (i.e.\ $d_{1,k}=\dots=d_{N,k}$).
On the other hand, if all the non-affine displacements are zero except for a single particle (e.g.\ $d_{1,k}>0$ and $d_{2,k}=\dots=d_{N,k}=0$),
we find $p_k=N^{-1}$ which goes to zero in the thermodynamic limit ($N\rightarrow\infty$).
Therefore, the participation ratio in the range $N^{-1}\leq p\leq 1$ can quantify the collectivity of the given displacement fields.
Note that we do not need to introduce any numerical thresholds to compute $p$ \cite{aval_quasi1}.

Figure \ref{fig:PDF_p} displays the probability distribution functions (PDFs) of the participation ratio, $P(p)$, for different shear rates.
Here the results with $N=16384$ are shown~\footnote{We also show $P(p)$ for different system sizes ($N=8192$, $16384$, and $32768$) in Appendix \ref{ap:Ndep}}.
If the shear rate is sufficiently large ($\dot{\gamma}t_0=10^{-3}$), the PDF exhibits a delta-peak around $p\simeq 0.5$.
Decreasing the shear rate, we observe that the peak position shifts to lower participation ratios and its height gets smaller (top panel).
If we further decrease the shear rate (bottom panel), the peak position continues to shift to lower $p$ and its height becomes much smaller (note the different scale of the vertical axes in Fig.\ \ref{fig:PDF_p}).
Interestingly, if the scaled shear rate is lowered to $\dot{\gamma}t_0 = 10^{-7}$, another peak emerges around $p\simeq 0.45$.
Such \emph{bimodal distributions} were not obtained from the quasi-static simulations \cite{aval_quasi1},
implying that two distinct types of \emph{characteristic events} are observed in systems sheared with finite shear rate ($\dot{\gamma}>0$).
This new emerging peak grows and shifts to higher $p$ with decreasing the shear rate in $\dot{\gamma}t_0\leq 10^{-7}$.
%
\subsection{Spatial structures of stress drop events}
\label{sec:spatial}
To gain an insight into characteristic stress drop events, we analyze their spatial structures.
Figure \ref{fig:vis}(a) shows an intensity plot of the PDFs, $P(p)$ (gray scale), where we combine all the data sets in Fig.\ \ref{fig:PDF_p} such that the scaled shear rate $\dot{\gamma}t_0$ varies along the horizontal axis (note that $P(p)$ is normalized for each $\dot{\gamma}$).
In this figure, the single peaks of the PDFs for large shear rates (Fig.\ \ref{fig:PDF_p}(top)) are represented by a ``ridge" in $\dot{\gamma}t_0 \gtrsim 10^{-7}$, while the two peaks for small shear rates (Fig.\ \ref{fig:PDF_p}(bottom)) are indicated by ``two ridges" in $\dot{\gamma}t_0 \lesssim 10^{-7}$.
If we denote the participation ratio at the ridge for large shear rates as $p_\mathrm{single}$ and those at the two ridges for small shear rates as $p_\mathrm{low}$ and $p_\mathrm{high}$ ($p_\mathrm{low}<p_\mathrm{high}$),
$p_\mathrm{single}$ is smoothly connected to $p_\mathrm{low}$ with decreasing the shear rate, whereas $p_\mathrm{high}$ suddenly appears around $\dot{\gamma}t_0 = 10^{-7}$.

Figures \ref{fig:vis}(b-g) display the non-affine displacements characterized by the participation ratios at the ridges, (b, c) $p_\mathrm{single}$, (d, f) $p_\mathrm{high}$, and (e, g) $p_\mathrm{low}$, where the sets of $p$ and $\dot{\gamma}t_0$ for (b-g) are indicated in Fig.\ \ref{fig:vis}(a).
If the shear rate is sufficiently large, the non-affine displacements are homogeneously distributed in space (Fig.\ \ref{fig:vis}(b)).
Even though their spatial structure seems to be random, the value $p_\mathrm{single}\simeq 0.5$ implies that their distribution is non-Gaussian
\footnote{The participation ratio is the reciprocal of kurtosis, where $p=1/3$ in the case of normal (Gaussian) distributions.}.
Decreasing the shear rate, we observe that the structure of non-affine displacements becomes heterogeneous (Fig.\ \ref{fig:vis}(c)).
If the system is driven by a very slow shear ($\dot{\gamma}\le 10^{-7}$), we find two types of structures;
large scale collective motions (Fig.\ \ref{fig:vis}(d)) and localized motions (Fig.\ \ref{fig:vis}(e)).
These two types of structures characterized by $p_\mathrm{high}$ and $p_\mathrm{low}$ are pronounced if we further decrease the shear rate (Figs.\ \ref{fig:vis}(f, g)).
Note that the collective motions span the whole system (Fig.\ \ref{fig:vis}(f)).
In addition, even if the participation ratios of two events are very close, the spatial structures can be different,\
e.g.\ the non-affine displacements in Figs.\ \ref{fig:vis}(b) where $p=0.443$ and (f) where $p=0.465$.
Therefore, the participation ratio $p$ alone is insufficient to describe the spatial structures of displacement fields during stress drop events.
%
%

%
%
\subsection{Coherent clusters}
\label{sec:corre_len}
To quantify the spatial expansion of non-affine displacements (Fig.\ \ref{fig:vis}(b-g)), we introduce \emph{coherent clusters}.
The participation ratio $p$ defined as Eq.\ (\ref{eq:p}) represents the fraction of \emph{mobile particles} \cite{aval_quasi1}.
According to Ref.\ \cite{aval_quasi1}, we first define mobile particles as those whose $d_i$ is in the top $p$ of the population.
For each stress drop event (Eq.\ (\ref{eq:drop})), we extract the mobile particles and divide them into coherent clusters.
Here, every constituent of a single coherent cluster is connected by a \emph{contact network}
(such that the mobile particle $i$ is a member of a coherent cluster if it makes a contact with any other mobile particle $j$ in the same cluster).
Then, we count the number of constituents of the largest coherent cluster as $N_{{\rm c},k}$, where $k$ stands for the event index.
Finally, we define the characteristic length of the $k$th event as
\begin{equation}
    \xi_k = d_0N_{{\rm c},k}^{1/d}
    \label{eq:xi}
\end{equation}
with the mean particle diameter $d_0$ and spatial dimensions $d=2$.
In Appendix \ref{ap:clusters}, we demonstrate how to extract the largest coherent cluster from the system for several values of $p$ and $\dot{\gamma}$.

\begin{figure}
\includegraphics[width=\linewidth]{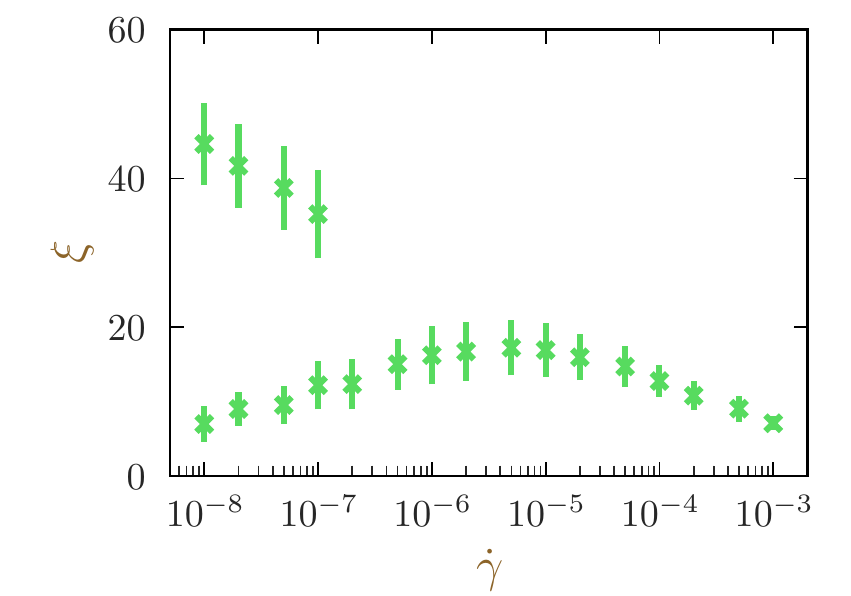}
\caption{
The size of the largest coherent cluster $\xi(p)$ (Eq.\ (\ref{eq:xi})) for the characteristic participation ratios,\
i.e.\ $\xi_\mathrm{single}$ for large shear rates ($\dot{\gamma}t_0\gtrsim 10^{-7}$),
and $\xi_\mathrm{low}<\xi_\mathrm{high}$ for small shear rates ($\dot{\gamma}t_0\lesssim 10^{-7}$).
The error bars represent standard deviations of $\xi(p)$ around $p=p_\mathrm{single}$, $p_\mathrm{high}$, and $p_\mathrm{low}$.
The system size is $N=16384$.
\label{fig:xi}}
\end{figure}

We analyze the average of $\xi_k$ over events at the peak position $p\approx p_{\alpha}\ (\alpha\in\{{\text{single},\text{high},\text{low}}\})$~\footnote{An event is regarded to belong to the peak at $p_\alpha$ if $p\in [p_\alpha^{\rm small},p_\alpha^{\rm large}]$ is satisfied, where $p_\alpha^{\rm small}$ and $p_\alpha^{\rm large}$ are the smallest/largest value of $p$ above/below which $P(p)>0.9P(p_\alpha)$ is satisfied.}.
The obtained characteristic sizes are denoted as $\xi_\mathrm{single}\equiv\xi(p_\mathrm{single})$, $\xi_\mathrm{high}\equiv\xi(p_\mathrm{high})$, and $\xi_\mathrm{low}\equiv\xi(p_\mathrm{low})$.
Figure \ref{fig:xi} shows the characteristic sizes 
as functions of the scaled shear rate $\dot{\gamma}t_0$, where the error bars indicate standard deviations.
In this figure, the lower (upper) branch represents $\xi_\mathrm{low}$ ($\xi_\mathrm{high}$)
so that $\xi_\mathrm{low}<\xi_\mathrm{high}$ holds in the low-rate regime ($\dot{\gamma}t_0\lesssim 10^{-7}$).
Decreasing the shear rate $\dot{\gamma}$, we find that (i) $\xi_\mathrm{single}$ for large shear rates continuously changes to $\xi_\mathrm{low}$, (ii) $\xi_\mathrm{high}$ appears around $\dot{\gamma}t_0=10^{-7}$, and (iii) $\xi_\mathrm{low}$ ($\xi_\mathrm{high}$) monotonously decreases (increases) in the low-rate regime.
Note that the trend of $\xi_\mathrm{single}$ in large shear rates is different from that of $p_\mathrm{single}$ which increases with the increase of $\dot{\gamma}$ (Fig.\ \ref{fig:vis}(a)), the reason of which will be explained in Appendix \ref{ap:clusters}.

\begin{figure}
\includegraphics[width=\linewidth]{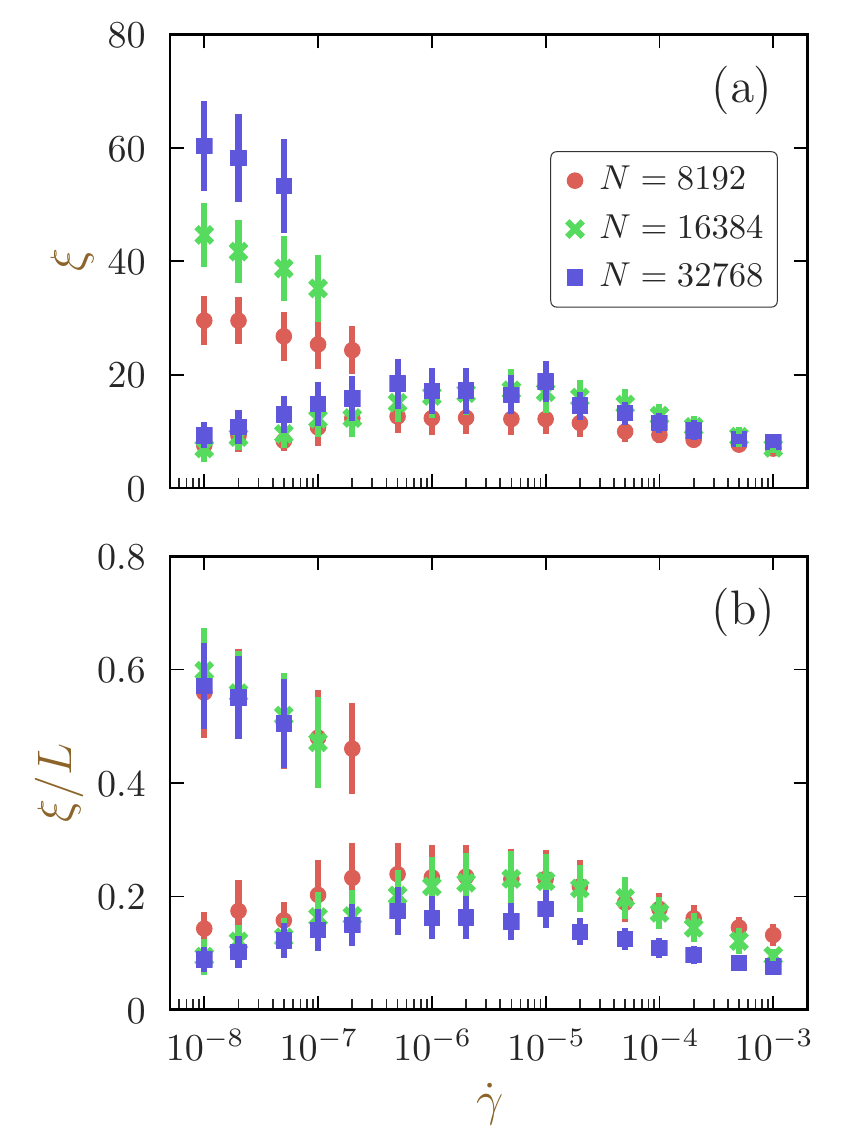}
\caption{
(a) The size $\xi(p)$ for the characteristic participation ratios,\ i.e.\ $\xi_\mathrm{single}$, $\xi_\mathrm{high}$, and $\xi_\mathrm{low}$,
where the system size increases from $N=8192$ to $32768$ (as listed in the legend).
(b) The characteristic sizes over the system length, $\xi/L$, where the symbols are as in (a).
The error bars are as in Fig.\ \ref{fig:xi}.
\label{fig:scaling_xi}}
\end{figure}

We further examine the system size dependence of the characteristic sizes, $\xi_\mathrm{single}$, $\xi_\mathrm{high}$, and $\xi_\mathrm{low}$.
Figure \ref{fig:scaling_xi}(a) displays the characteristic sizes as functions of the scaled shear rate $\dot{\gamma}t_0$,
where different symbols are the results of different system sizes $N$ (as listed in the legend).
As can be seen, $\xi_\mathrm{high}$ (upper branch) monotonously increases with the increase of $N$, while $\xi_\mathrm{single}$ for large shear rates is quite insensitive to the system size.
The size $\xi_\mathrm{low}$ (lower branch) for small shear rates slightly increases with $N$
although it becomes almost independent of the system size in the zero shear rate limit, $\dot{\gamma}\rightarrow 0$.
As shown in Fig.\ \ref{fig:scaling_xi}(b), all the data of $\xi_\mathrm{high}/L$ are nicely collapsed on top of each other,
meaning that the upper branch is proportional to the linear dimension of the system,\ i.e.\
\begin{equation}
\xi_\mathrm{high}\propto L~.
    \label{eq:xi_high_L}
\end{equation}
Therefore, the large scale collective motions, which extend over the system (Fig.\ \ref{fig:vis}(d, f)), linearly scale with $L$.
On the other hand, the localized motions (Fig.\ \ref{fig:vis}(e, g)) are not affected by the system size.

%
\subsection{Time evolution of participation ratio}\label{sec:t_e}
\begin{figure*}
\includegraphics[width=\linewidth]{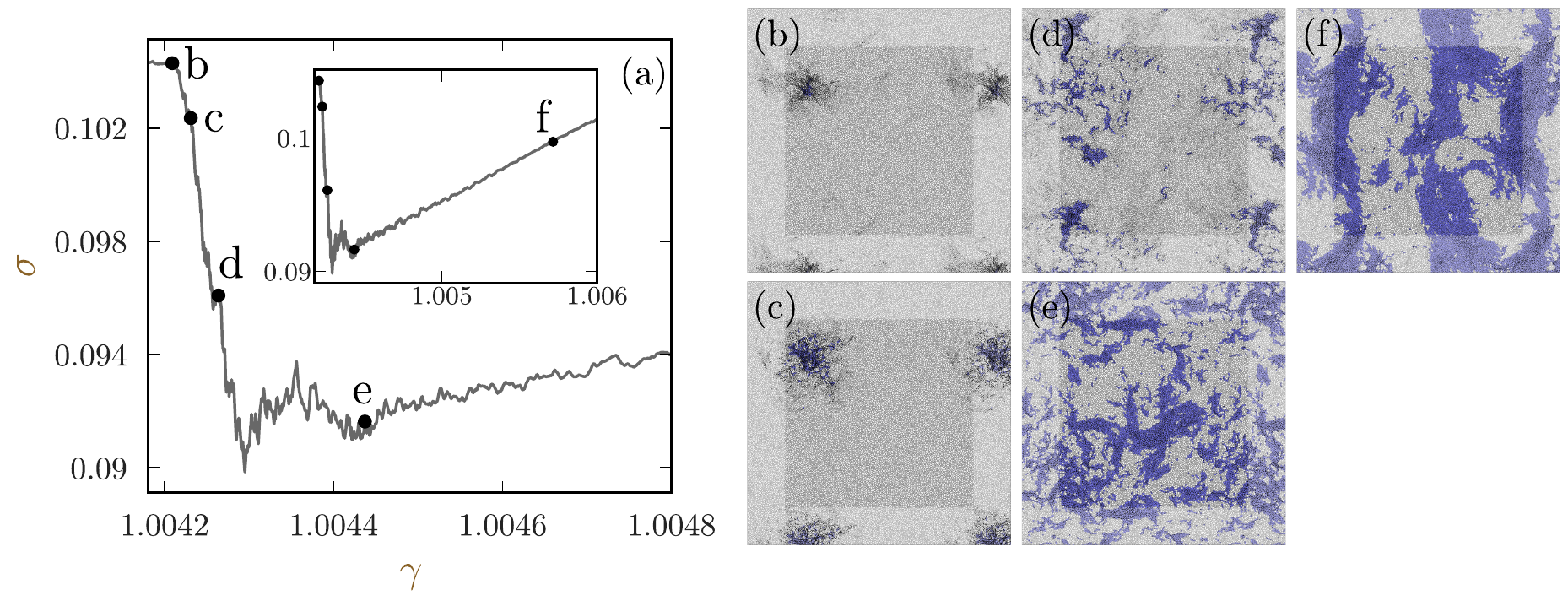}
\caption{
  (a): Close-up stress-strain curve to a large stress drop event chain for
  the system with $N=16384$ and $\dot{\gamma}=1\times 10^{-8}$.
  (Inset): The same data with a wider range of the horizontal axis.
  (b-f): Visualization of displacement fields during events marked
  in (a).
  The dots in (a) which represents the position of events (b-f) are
  put at the beginning of events.
  See the caption of Fig.~\ref{fig:vis} for the details of the visualization.
  \label{fig:te1}}
\end{figure*}

Thus far, we have analyzed the spatial structures of nonaffine displacement fields during stress drop events.
In particular, we showed that qualitatively different two types of characteristic stress drop events are observed under very low shear rates and clarified that the events at $p\approx p_{\rm high}$ peak are system-spanning while those at $p\approx p_{\rm low}$ are localized.
In this section, to elucidate the physical origin of these two characteristic events, we examine the time evolution of events.
Under AQS shear, the stress-strain curves are divided into two distinct branches~\cite{Saitoh2019SM}: the elastic branch where the stress increases linearly with the applied strain and the plastic branch where the stress drops abruptly.
Under finite-rate shear, different behaviors show up as explained below.

\subsubsection{Low-rate regime}
We show a close-up plot of the stress-strain curve around a large
stress drop for the system
with $\dot{\gamma}=1\times 10^{-8}$ in Fig.~\ref{fig:te1}(a).
Under finite-rate shear, even with this slowest value, we observe qualitative differences from the results for systems under AQS shear: the stress-strain curve exhibits small oscillations during and after a large stress drop event that is expected to correspond to a stress drop event under AQS shear.

The chain of events composing this large stress drop event seems to originate from two main events.
The displacement field during the first one is visualized in Fig.~\ref{fig:te1}(b).
Here, we see a clear quadrupolar pattern, indicating that this event
is the primary ST.
The energy released from this ST propagates throughout the system via the elastic field.
Interestingly, this propagation process is accompanied by small
oscillations of stress, leading to multi-stage stress drops.
The displacement field during one of them is visualized in
Fig.~\ref{fig:te1}(c):
we observe a misty pattern around the primary ST.
Ref.~\cite{Oyama2020INM} has conducted the instantaneous normal mode
analysis for sheared glasses  and reported that most eigenvalues are positive even under finite rate shear (and STs, the signatures of plasticity, correspond to negative eigenvalues).
According to this knowledge, we speculate that the oscillations during
sharp stress drops are the consequence of the excitation of the normal
modes by the energy released from the primary ST.
Since those modes are elastic, resulting stress response is oscillatory.

The released energy propagates throughout the system and finally triggers the second main event.
As visualized in Fig.~\ref{fig:te1}(d), this event is another quadrupolar ST that is adjacent to the primary one (see left bottom part of the original cell; the rest are the residue of the primary ST).
Therefore, this event is likely to be the secondary ST induced by the released energy from the primary one: this is a typical formation of an avalanche~\cite{MaloneyPRE2006}.
Importantly, although the events shown in Fig.~\ref{fig:te1}(b) where $p=0.003$ and (c) where $p=0.021$ belong to the low-$p$ peak in $P(p)$ ($p\in[0,0.023]$ with $p_{\rm low}=0.013$), the secondary event (Fig.~\ref{fig:te1}(d)) has a slightly larger value of $p=0.076$ because of the residue of the primary ST.
In other words, the low-$p$ peak is composed of events with single ST.
We again stress that the peak includes small oscillatory events during large stress drop events.

After a sharp and large stress drop event due to an avalanche (a chain of STs), there appears another type of oscillation of the stress.
While the stress oscillations mentioned in the previous paragraphs take place during a large stress event where the global trend of the stress is decreasing, oscillations that we are now discussing occur after such a decreasing trend.
Instead, these oscillations are observed in the elastic-like branch where the global trend of the stress is increasing (Figs.~\ref{fig:te1}(e,f)).
From the view point of the instantaneous normal modes, they can be again considered to be the consequence of the excitation of normal modes due to the released energy from the avalanche.
In this sense, we name these events {\it posterior-excitation events}.
Regarding the precise spatial pattern of them, these events exhibit system-spanning coherent motions (Fig.~\ref{fig:te1}(e)).
This may be because short-wave-length modes are damped fast during dissipative dynamics~\cite{Hansen} and long-wave-length modes survive longer.
These events indeed compose the peak at $p\approx p_{\rm high}$.
Although the longest-wave-length phonon modes are manifested eventually as shown in Fig.~\ref{fig:te1}(f), they have higher values of $p=0.514$ than $p_{\rm high}=0.443$.

To summarize, under finite rate shear, even in the very low rate limit, the stress-strain curve exhibits qualitative difference from the one under AQS shear.
Concretely, an avalanche event is divided into sub-drop events
corresponding to single STs.
Moreover, each ST event is further composed of multi-stage stress drops (or oscillatory behavior).
Among them, events related to a single ST, including transient
oscillatory motions, form the low-$p$ peak of $P(p)$.
After a large stress drop, there appear another type of stress oscillation, which we call posterior-excitations.
Although those oscillations ultimately become the longest-phonon-mode-like coherent motion,
the high-$p$ peak of $P(p)$ corresponds to rather transient complicated motions such as the one visualized in Fig.~\ref{fig:te1}(e).
\begin{figure*}
\includegraphics[width=0.655\linewidth]{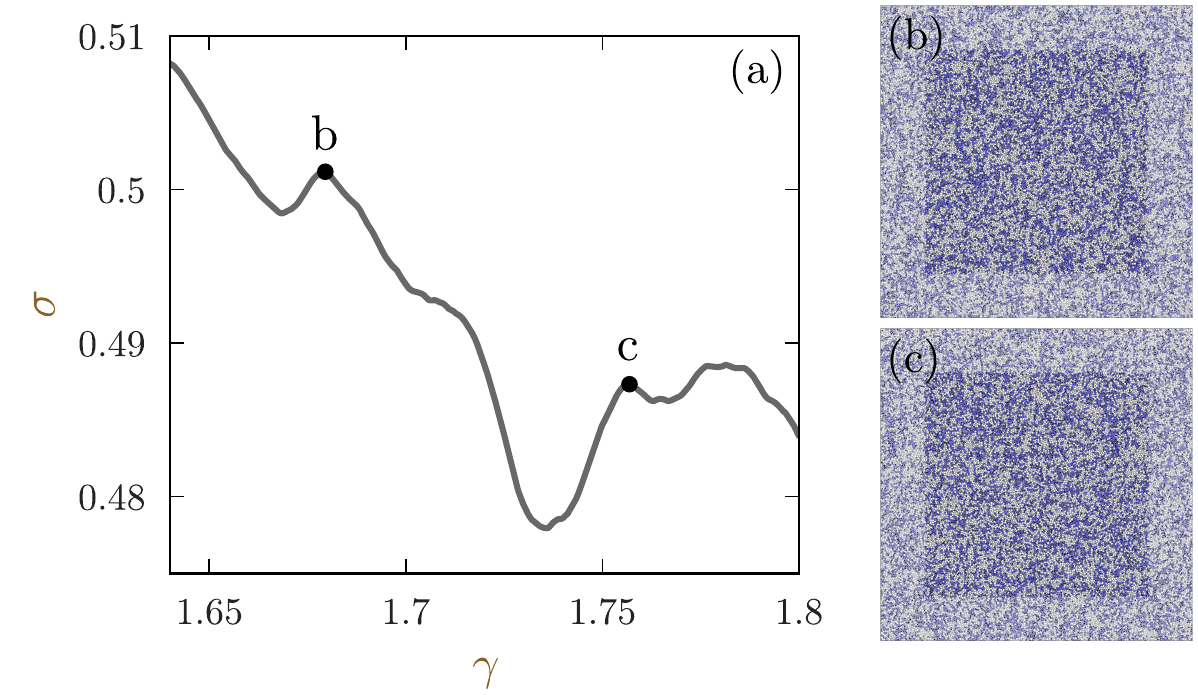}
\caption{
  (a) Stress-strain curve for
  the system with $N=16384$ and $\dot{\gamma}=1\times 10^{-5}$.
  (b,c) Visualization of displacement fields during events marked
  in (a).
  The dots in (a) which represents the position of events (b,c) are
  put at the beginning of events.
  See the caption of Fig.~\ref{fig:vis} for the details of the visualization.
\label{fig:te2}}
\end{figure*}

\subsubsection{High-rate regime}
We next present the results for the high-rate regime.
In Fig.~\ref{fig:te2}(a), we plot a close-up plot of the stress-strain curve for the system under the highest shear rate $\dot{\gamma}=1\times 10^{-3}$.
Under such a high rate, the fluctuations in the stress become smooth and it is impossible to distinguish between plastic and elastic parts.
Probably, this is because multiple STs (or even avalanches) are always taking place simultaneously and the effects of them are averaged out~\cite{Oyama2020INM}.
Therefore, a single "stress drop event" includes contributions from
multiple plastic events
and $P(p)$ becomes unimodal with a peak that represents random displacement fields (Fig.~\ref{fig:te2}(b, c)).
We stress that this random displacement filed possesses non-Gaussianity as mentioned in \ref{sec:bimodal}.

\section{Discussion}
\label{sec:discussion}
In this study, we have numerically investigated microscopic insights into plastic deformations of soft athermal particles sheared with finite shear rate, $\dot{\gamma}>0$.
We focused on the stress drop events (Eq.\ (\ref{eq:drop})) in a steady state and computed the participation ratios $p$ (Eq.\ (\ref{eq:p})) of non-affine displacements of the particles.
The participation ratio in the range $N^{-1}\leq p\leq 1$ quantifies how localized the non-affine displacements are,
where we found that the PDF of participation ratios $P(p)$ is strongly dependent on the shear rate:
The PDF only has a single peak if the shear rate is large, whereas it becomes a bimodal distribution in the low-rate regime (Fig.\ \ref{fig:PDF_p}).
The peak positions of the PDFs,\ i.e.\ $p_\mathrm{single}$, $p_\mathrm{high}$, and $p_\mathrm{low}$, specify the types of non-affine displacements
as (i) homogeneous distributions in space, (ii) large scale collective motions, and (iii) spatially localized structures, respectively (Fig.\ \ref{fig:vis}).
By using the participation ratios, we have extracted mobile particles from the system to measure the ensemble-averaged size of the largest coherent cluster, $\xi(p)$ (Eq.\ (\ref{eq:xi})).
The characteristic sizes, $\xi_\mathrm{single}$, $\xi_\mathrm{high}$, and $\xi_\mathrm{low}$ (corresponding to the characteristic participation ratios, $p_\mathrm{single}$, $p_\mathrm{high}$, and $p_\mathrm{low}$, respectively), exhibit the dependence on the shear rate $\dot{\gamma}$ which is similar to that of the participation ratios (Fig.\ \ref{fig:xi}).
Note that, however, their trends in large shear rates are different, where $\xi_\mathrm{single}$ ($p_\mathrm{single}$) decreases (increases) with increasing $\dot{\gamma}$.
We have clarified that the size $\xi_\mathrm{high}$ linearly scales as $\xi_\mathrm{high}\sim L$ with the system length $L$ (Eq.\ (\ref{eq:xi_high_L})),
indicating that the large scale collective motions extend over the system \cite{pdf1}.
On the other hand, the size $\xi_\mathrm{low}$ is insensitive to $L$ and becomes independent of the system size in the zero shear limit, $\dot{\gamma}\rightarrow 0$ (Fig.\ \ref{fig:scaling_xi}).
This means that the non-affine displacements during the stress drop event are localized as in the case of AQS simulations.

In the following, we discuss previous results of correlation lengths in soft athermal particles under shear which are associated with $\xi_\mathrm{high}$ (Sec.\ \ref{sub:rigid}) and $\xi_\mathrm{low}$ (Sec.\ \ref{sub:aval}).
%
%
%
\subsection{Rigid clusters}
\label{sub:rigid}
The spatial correlation function of the instantaneous non-affine velocity field $C_v(r)$ is often utilized to define a dynamical correlation length in sheared amorphous solids.
Specifically, the distance at which $C_v(r)$ takes the minimum or the decay length obtained by the fitting to the exponential decaying function have been employed as the correlation length in this context.
In these definitions, the statistical averages are simply calculated over randomly sampled configurations without taking into consideration whether the system is in the elastic branch or the plastic branch (or the posterior-excitation events).
Many works have studied such correlation lengths in sheared dense soft athermal disks and it has been established that the correlation length based on this definition span the whole system if the shear rate is slow enough.

As shown in Fig.~\ref{fig:te1}(a), if we randomly sample configurations, it is very probable to obtain states in the elastic branch or the posterior-excitation events.
This is one major difference from our $\xi_{\rm low}$ which is
measured during the plastic branch.
Also, although our $\xi_{\rm low}$ is calculated based only on the magnitude of
displacement vectors during one whole stress drop event, 
$C_v(r)$ takes the vectorial information into consideration and
utilizes only the instantaneous information.

\subsection{Avalanche correlation length}
\label{sub:aval}
As an example of the correlation length defined based on plastic events, the one of avalanches, which we call $\xi_{\rm ava}$, has been discussed in several contexts~\cite{LemaitrePRL2009,LinPNAS2014,Oyama2020INM}.
Here we discuss the relation between $\xi_{\rm ava}$ and our $\xi_{\rm low}$, which directly reflect the dynamics due to the plastic events (or, STs).

Since we want to make a comparison with $\xi_{\rm low}$ that is defined only in the low-rate limit, we are interested in the feature of $\xi_{\rm ava}$ in the limit of $\dot{\gamma}\to 0$.
The length scale $\xi_{\rm ava}$ is known to diverge and span the whole system in the low rate limit (when $\dot{\gamma}$ increases, multiple avalanches start taking place simultaneously and $\xi_{\rm ava}$ starts decreasing~\cite{LinPNAS2014, Oyama2020INM}).
This length $\xi_{\rm ava}$ has been introduced based on the yielding criticality~\cite{LinPNAS2014, Oyama2020INM} and can
be associated with the so-called cutoff avalanche size $S_c$ through
the relation $S_c\sim L^{d_f}\sim \xi_{\rm ava}^{d_f}$, which becomes well-defined under AQS shear.
Roughly speaking, $S_c$ measured under AQS shear corresponds to the largest event for a given system size and is composed of a system-spanning avalanche of STs~\cite{Oyama2020INM}.

On the other hand, our $\xi_{\rm low}$ provides the size of typical events, not the largest one: such a size corresponds to that of a single ST.
System-spanning avalanches are known to take place more rarely and have much higher participation ratio than single STs~\cite{Oyama2020avalanche, Oyama2020INM}.
{Correspondingly, system-spanning ones do not belong to the peak around $p\approx p_{\rm low}$.}
This means that $\xi_{\rm low}$ and $\xi_{\rm ava}$ are intrinsically
different length scales in nature.

\subsection{Future works}
\label{sec:summary}
In our numerical simulations, we fixed the packing fraction of the particles to $\phi=0.88$, which is far above the jamming transition density $\phi_J$ \cite{gn3,gn4}, such that the system is in a solid phase.
If the system is static (and \emph{above} jamming), it is known that several length scales associated with
e.g.\ elastic heterogeneities \cite{local_e3,tanguy0,tanguy1,tanguy2,tanguy3}, continuum limits of disordered systems \cite{tanguy0,tanguy1,tanguy2,tanguy3,rs0,rs1}, and phonon transports \cite{phonon_transport} diverge as the system approaches the onset of unjamming.
On the other hand, if the system is \emph{below} jamming (is in a liquid phase), the correlation length of non-affine displacements exhibits critical divergence near the transition \cite{rheol0,corl2,corl3,saitoh14}.
Therefore, it is an important next step to investigate the dependence of the characteristics sizes, $\xi_\mathrm{high}$ and $\xi_\mathrm{low}$, on the proximity to jamming, $|\Delta\phi|=|\phi-\phi_J|$, as well as to clarify the critical behavior of coherent clusters.

In our MD simulations, we used the spring-dashpot model for the interaction between the particles in contact (Sec.\ \ref{sec:methods}) as a canonical model of frictionless granular materials \cite{dem}.
However, it is known that interaction forces drastically change the rate-dependent flow behavior of soft athermal particles,\
e.g.\ the force law of viscous damping, $\bm{f}_{ij}^\mathrm{vis}$, controls the shear thinning/thickening \cite{rheol7}, either the continuous \cite{CST0,CST1} or discontinuous shear thickening of dry granular materials \cite{DST0,DST2,DST5,DST6} and colloidal suspensions \cite{DST1,DST4,DST7,DST8} is induced by the microscopic friction, and the flow curves are nonmonotonic if one introduces cohesive forces \cite{AST0,AST1}.
Therefore, it is interesting to examine how the interaction forces affect the large scale collective motions and localized structures of non-affine displacements.


As another future direction, it is an interesting question whether we observe similar bimodal shape of $P(p)$ in systems under different forms of perturbations(such as air~\cite{DurianNatPhys}, vibration~\cite{vibration}, compression~\cite{Antonaglia2014SR}/decomression~\cite{Shimada2020} or active noises~\cite{FilyPRL2012, BerthierNJP2017, OyamaPRR2019}).
\begin{acknowledgments}
We thank H. Hayakawa, M. Imai, and T. Kawakatsu for fruitful discussions.
This work was financially supported by JSPS KAKENHI Grant Numbers 18K13464, 20H01868, 20J00802, and 20K14436.
\end{acknowledgments}
\begin{appendix}
\setcounter{figure}{0}
\renewcommand{\thefigure}{A\arabic{figure}}
\section{PDFs of participation ratios for different system sizes}
\label{ap:Ndep}
\begin{figure*}
  \includegraphics[width=\linewidth]{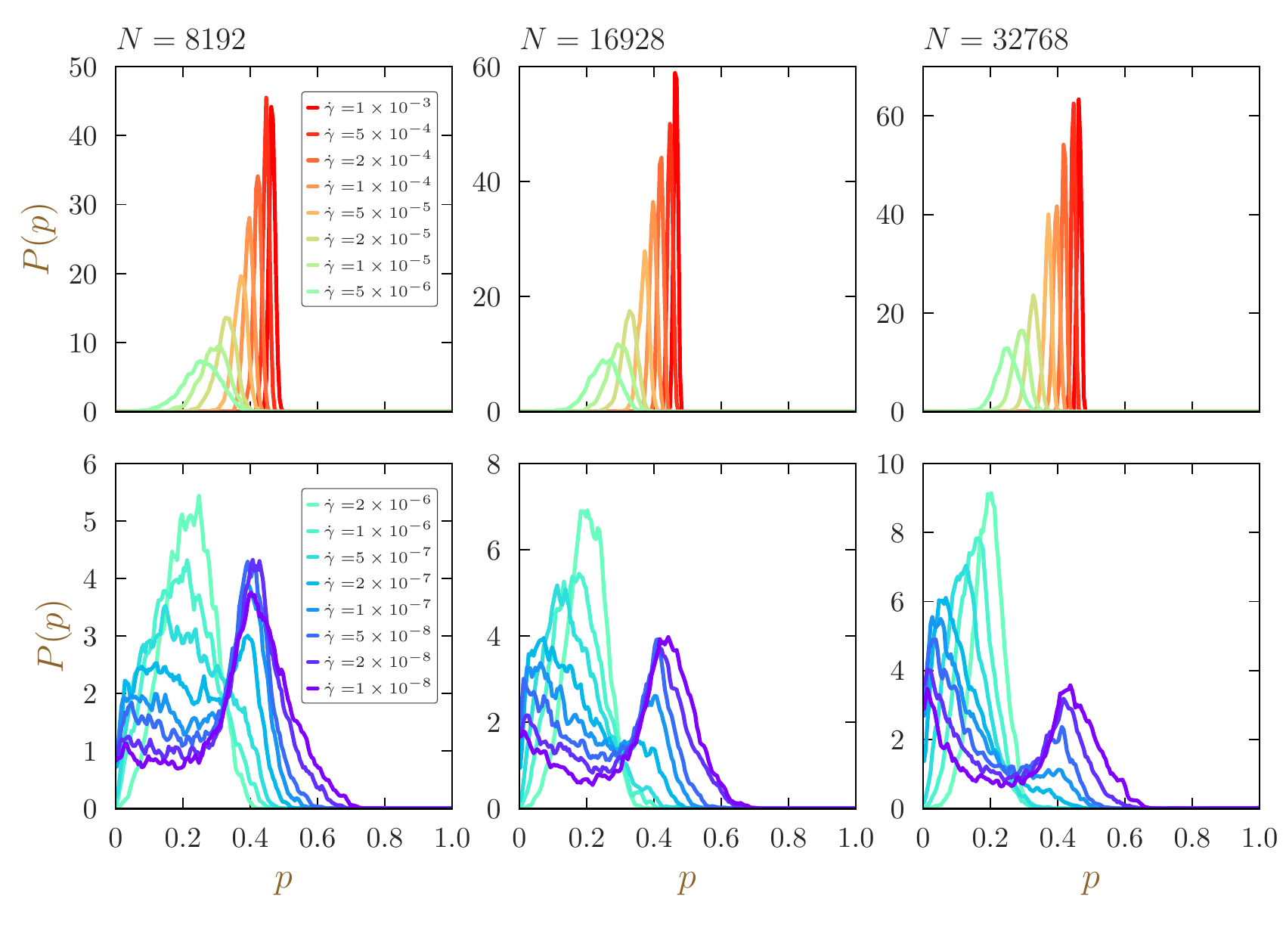}
  \caption{
    PDFs of participation ratios $P(p)$,
    where the system size increases as $N=8192$, $16384$, and $32768$ from left to right.
    The results for high shear rates (top) and low shear rates (bottom) are displayed.
    Different colors represent different values of the scaled shear rate $\dot{\gamma}t_0$ as listed in the legends in the left column.
\label{fig:PDF_p_other}}
\end{figure*}

In this appendix, we examine the dependence of the PDFs of participation ratios, $P(p)$, on the system size $N$ and shear rate $\dot{\gamma}$.
Figure \ref{fig:PDF_p_other} displays the PDFs for different system sizes, $N=8192$, $16384$, and $32768$ (from left to right), where the scaled shear rate decreases from $\dot{\gamma}t_0=10^{-3}$ to $10^{-8}$ (from top to bottom) as listed in the legends.
If the shear rate is large (top panel), the PDFs have single peaks of which heights (widths) become high (narrow) with the increase of $N$ (note that ranges of the vertical axis are different).
On the other hand, if the shear rate is small (bottom panel), the PDFs exhibit characteristic two peaks,\ i.e.\ low- and high-$p$ peaks.
As in the case of AQS simulations \cite{aval_quasi1}, the low-$p$ peak slightly shifts to small participation ratios with increasing the system size $N$.
The heights of the low-$p$ peaks increase with the increase of $N$, while those of the high-$p$ peaks monotonously decrease with increasing $N$.
%
\section{Coherent clusters}
\label{ap:clusters}
In this appendix, we explain how to extract \emph{coherent clusters} from the system for each stress drop event.

\begin{figure*}
    \includegraphics[width=\linewidth]{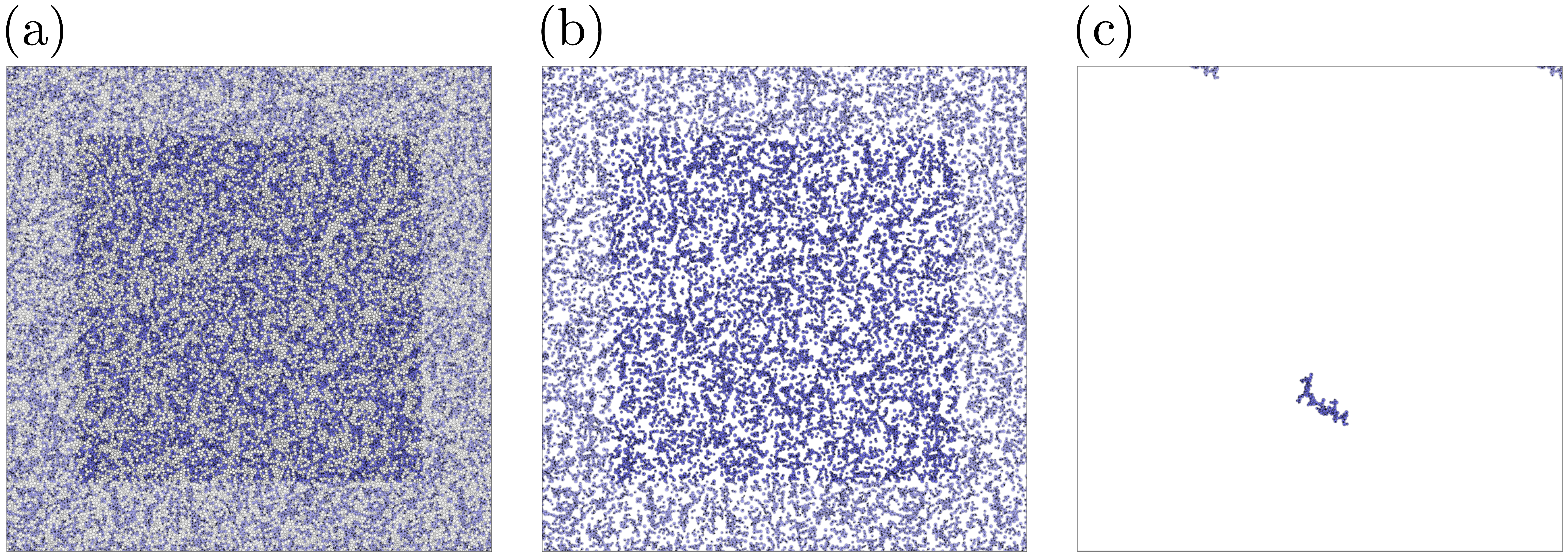}
    \caption{
    (a) Non-affine displacements (arrows) during a stress drop event, where we use $N=16384$ and $\dot{\gamma}t_0=10^{-3}$.
    The mobile (blue) and immobile (white) particles are distinguished by color.
    (b) The non-affine displacements (arrows) of the mobile (blue) particles.
    (c) The largest coherent cluster extracted from the system.
    Here, the participation ratio is $p_\mathrm{single}=0.467$ (located at the single peak of the PDF $P(p)$, see Fig.\ \ref{fig:PDF_p_other})
    and the characteristic size is given by $\xi_\mathrm{single}=0.089L$.
    \label{fig:coherent_clusters3}}
  \end{figure*}

Figure \ref{fig:coherent_clusters3}(a) shows non-affine displacements (arrows) during a stress drop event, where the system is deformed with a large shear rate, $\dot{\gamma}t_0=10^{-3}$.
As shown in Fig.\ \ref{fig:coherent_clusters3}(b),
we extract \emph{mobile particles} (blue) from the system and divide them into coherent clusters (see Sec.\ \ref{sec:corre_len} for the definitions).
Then, we find the largest coherent cluster as displayed in Fig.\ \ref{fig:coherent_clusters3}(c).
The participation ratio of this stress drop event is $p_\mathrm{single}=0.467$, where the size of the largest coherent cluster is estimated as $\xi_\mathrm{single}=0.089L$.

\begin{figure*}
    \includegraphics[width=\linewidth]{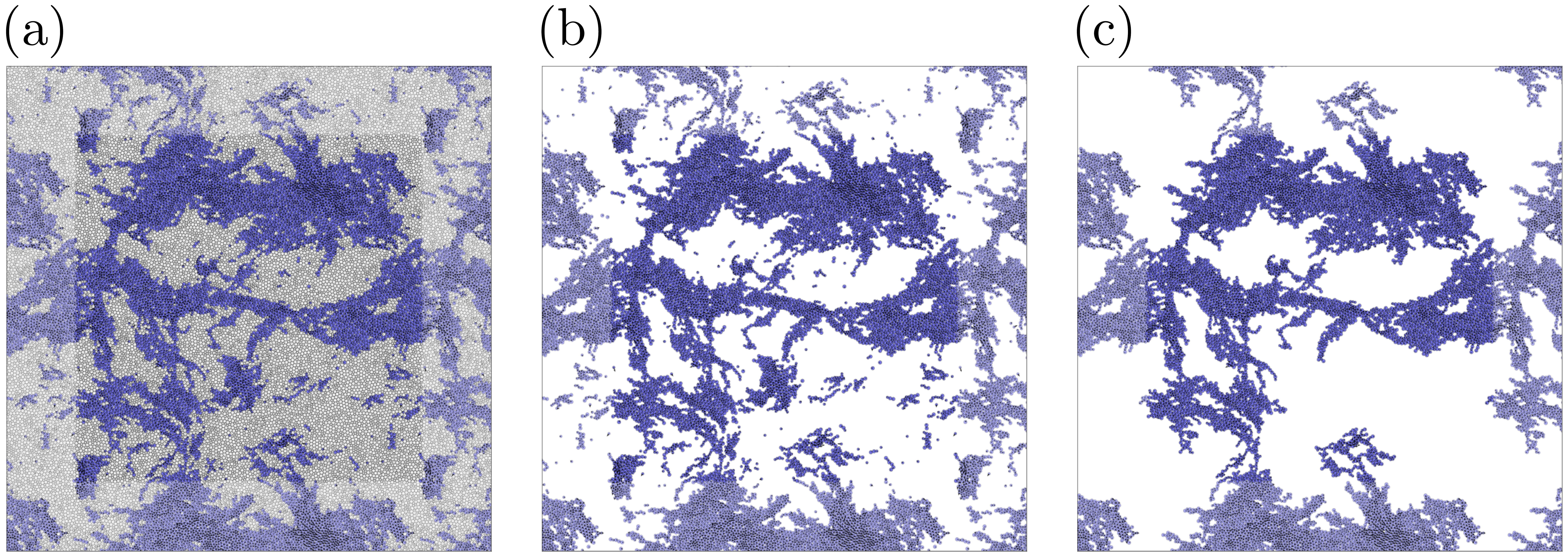}
    \caption{
    (a) Non-affine displacements (arrows) during a stress drop event, where we use $N=16384$ and $\dot{\gamma}t_0=10^{-8}$.
    The mobile (blue) and immobile (white) particles are distinguished by color.
    (b) The non-affine displacements (arrows) of the mobile (blue) particles.
    (c) The largest coherent cluster extracted from the system.
    Here, the participation ratio is $p_\mathrm{high}=0.457$ (located at the high-$p$ peak of the PDF, see Fig.\ \ref{fig:PDF_p_other})
    and the characteristic size is given by $\xi_\mathrm{high}=0.667L$.
    \label{fig:coherent_clusters1}}
  \end{figure*}

\begin{figure*}
    \includegraphics[width=\linewidth]{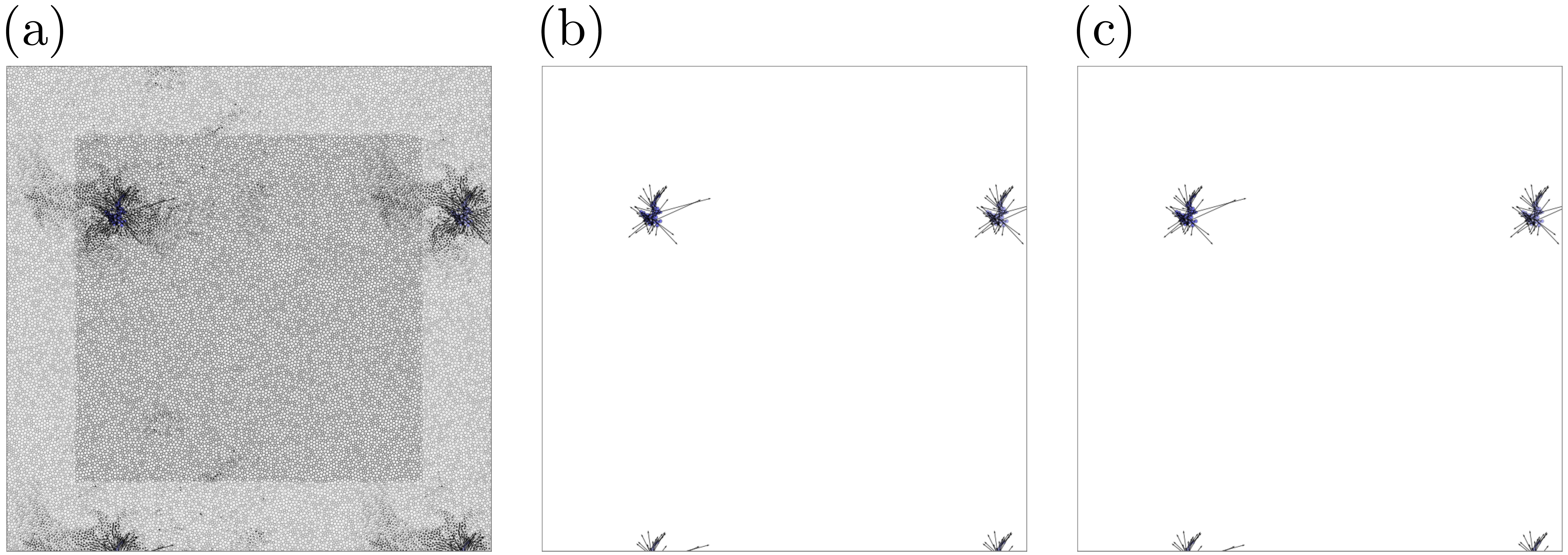}
    \caption{
    (a) Non-affine displacements (arrows) during a stress drop event,
    where $N$, $\dot{\gamma}$, and the color are as in Fig.\ \ref{fig:coherent_clusters1}.
    (b) The non-affine displacements (arrows) of the mobile (blue) particles.
    (c) The largest coherent cluster extracted from the system.
    Here, the participation ratio is $p_\mathrm{low}=0.003$ (located at the low-$p$ peak of the PDF, see Fig.\ \ref{fig:PDF_p_other})
    and the characteristic size is given by $\xi_\mathrm{low}=0.053L$.
    \label{fig:coherent_clusters2}}
  \end{figure*}

Figures \ref{fig:coherent_clusters1}(a) and \ref{fig:coherent_clusters2}(a) also show non-affine displacements (arrows) during stress drop events, where the system is deformed with a small shear rate, $\dot{\gamma}t_0=10^{-8}$.
We extract mobile particles from the system (Figs.\ \ref{fig:coherent_clusters1}(b) and \ref{fig:coherent_clusters2}(b))
and find the largest coherent clusters (Figs.\ \ref{fig:coherent_clusters1}(c) and \ref{fig:coherent_clusters2}(c)).
In Fig.\ \ref{fig:coherent_clusters1}, the participation ratio and size of the largest coherent cluster are given by $p_\mathrm{high}=0.457$ and $\xi_\mathrm{high}=0.667L$, respectively.
On the other hand, those in Fig.\ \ref{fig:coherent_clusters2} are $p_\mathrm{low}=0.003$ and $\xi_\mathrm{low}=0.053L$.

Even though the participation ratios in Figs.\ \ref{fig:coherent_clusters3} and \ref{fig:coherent_clusters1} are nearly equal, $p_\mathrm{single}\simeq p_\mathrm{high}$, the characteristic sizes are very different, $\xi_\mathrm{single}\ll\xi_\mathrm{high}$.
Accordingly, the spatial structures of non-affine displacements are also different (Figs.\ \ref{fig:coherent_clusters3} and \ref{fig:coherent_clusters1}).
Similarly, the characteristic sizes in Figs.\ \ref{fig:coherent_clusters3} and \ref{fig:coherent_clusters2} are close to each other, $\xi_\mathrm{single}\simeq \xi_\mathrm{low}$, though the participation ratios are different, $p_\mathrm{single}\gg p_\mathrm{low}$, and the spatial structures are totally different (Figs.\ \ref{fig:coherent_clusters3} and \ref{fig:coherent_clusters2}).
Therefore, the structures of non-affine displacements cannot be quantified only by the participation ratio $p$ or the characteristic size $\xi$.
%
%
%
%
\end{appendix}
\bibliography{two_length}
\end{document}